\documentclass[reprint,amsmath,amssymb,aps,floatfix,showkeys,showpacs,
]{revtex4-1}

\usepackage[T1]{fontenc}
\usepackage[utf8]{inputenc}

\usepackage{graphicx,colordvi}
\usepackage{dcolumn}
\usepackage{bm}

\begin{document}


\title{Shallow water soliton dynamics beyond KdV}
\thanks{One of the authors (A.K.) thanks for support from "Stochastic Analysis Research Network", grant PIA-CONICYT-Scientific Research Ring \#1112.
}

\author{Anna Karczewska}
 \email{A.Karczewska@wmie.uz.zgora.pl}
\affiliation{Faculty of Mathematics, Computer Science and Econometrics\\ University of Zielona G\'ora, Szafrana 4a, 65-246 Zielona G\'ora, Poland}

\author{Piotr Rozmej}
 \email{P.Rozmej@if.uz.zgora.pl}
\affiliation{Institute of Physics, Faculty of Physics and Astronomy \\
University of Zielona G\'ora, Szafrana 4a, 65-246 Zielona G\'ora, Poland}
\date{\today} 

\author{Eryk Infeld} \email{Eryk.Infeld@ncbj.gov.pl}
\affiliation{National Centre for Nuclear Research, Hoża 69, 00-681 Warszawa, Poland}

\begin{abstract}
An alternative way for the derivation of the new KdV-type equation is presented. The equation contains terms depending on the bottom topography (there are six new terms in all, three of which are caused by the unevenness of the bottom). It is obtained in the second order perturbative approach in the weakly nonlinear, dispersive and long wavelength limit. Only treating all these terms in the second order perturbation theory made the derivation of this KdV-type equation possible. The motion of a wave, which starts as a KdV soliton, is studied according to the new equation in several cases by numerical simulations. The quantitative changes of a soliton's velocity and amplitude appear to be directly related to bottom variations. Changes of the soliton's velocity appear to be almost linearly anticorrelated with changes of water depth whereas correlation of variation of soliton's amplitude with changes of water depth looks less linear. When the bottom is flat, the new terms narrow down the family of exact solutions, but at least one single soliton survives. This is also checked by numerics.

\end{abstract}



\maketitle

\section{Introduction} \label{intro}

The ubiquitous Korteveg de Vries equation \cite{kdv} is a common approximation for several problems in nonlinear physics. One of these problems is the {\em shallow water wave problem} extensively studied during the last fifty years and described in many textbooks and monographs (see, e.g.\ \cite{Whit,DrJ,InR,Ablc,Rem,Hir}). 
The KdV equation corresponds to the case when the water depth is constant. There have been numerous attempts to study nonlinear waves in the case of a non-flat bottom. 
One of the first attempts to incorporate bottom topography is due to  Mei and  Le M\'ehaut\'e~\cite{Mei}. However, the authors did not obtain any simple KdV-type equation.
Among the first papers treating a slowly varying bottom is Grimshaw's paper \cite{Grim70}. He obtained an asymptotic solution describing a slowly varying solitary wave above a slowly varying bottom. For small amplitudes the wave amplitude varies inversely as the depth.  
Djordjevi\'c and Redekopp \cite{Djord} studied the development of packets of surface gravity waves moving over an uneven bottom. They derived the variable coefficient nonlinear Schr\"odin\-ger equation (NLS) for such waves and
using expansion in a single small parameter they found  fission of an envelope soliton. A similar approach was later developed by Benilow and Howlin. This fission from the NLS has been found in other physical contexts \cite{BH,InR}.

We point out papers \cite{Pel,Peli,Peli1} as  examples of approaches which combine  linear and nonlinear theories.
For instance, in \cite{Pel} the authors study long-wave scattering by piecewise-constant periodic topography for solitary-like wave pulses and for KdV solitons. 
Another extensively investigated approach is the Gardner equation  (sometimes called the forced KdV equation) \cite{Grim,Smy,Kam}. 
Unidirectional waves over a slowly varying bottom have been studied  
by Van Groeasen and Pudjaprasetya \cite{G&P1,G&P2} within a Hamiltonian approach. For a slowly varying bottom, they obtained a forced KdV-type equation. The discussion of that equation gives  an increase of the amplitude and decrease of the wavelength when a solitary wave enters a shallower region.
The Green-Naghdi equations follow when taking an appropriate average of vertical variables \cite{GN,Nad,Kim}.
 Another study of long wave propagation over a submerged 2-dimensional bump was recently presented in \cite{NiuYu}, albeit according to linear long-wave theory.

Recently, an interesting numerical study of solutions to the free-surface Euler equations in the conformal-mapping formulation has been published by the team working  within the MULTIWAVE project \cite{MULTI}. The authors illustrate that approach by numerical results for soliton fission over a submerged step and supercritical stream over a  submerged obstacle  \cite{CV_DD}.

In this paper we briefly summarize the derivation of a   KdV-type equation,  second order in small parameters, containing terms from the bottom function, derived recently by two of us and Rutkowski in \cite{KRR}. Next we present some examples of the evolution of a KdV soliton according to
that equation, obtained in numerical simulations, stressing changes of soliton's velocity and amplitude when the wave passes over an extended obstacle or hole.
It is worth noting that the equation derived in \cite{KRR} is a KdV-like equation of the second order, a single evolution equation for surface waves which contains terms for a bottom variation. In this context see a paper by Kichenassamy and Olver "Existence and nonexistence of solitary wave solutions to higher-order model evolution equations" \cite{SK_PO}. The authors claimed {\it for most of higher-order models, but only those which reduce to KdV solitary waves in an appropriate scaling limit, solitary wave solutions of the appropiate form do not exist!}
On the other hand Burde \cite{Burde} presents solitary wave solutions of the higher-order KdV models for bi-directional water waves.

The paper is organized as follows. In section \ref{probl} the shallow water problem is set and expressed in non-dimen\-sional variables. Section \ref{deriv} contains the derivation of the second order wave equation sligthly different from that presented in the previous paper \cite{KRR}. The existence of at least one conservation law is proved. In section \ref{nsol} an analytic solution to the second order KdV-type equation with an even bottom is found. The solution has the single-soliton form. The possible existence of multi-soliton solutions for that equation is still an open question. Section \ref{numer} presents several cases of time evolution of the KdV soliton governed by the second order KdV-type equation with terms from an uneven bottom obtained in numerical simulations.

\section{Problem setting} \label{probl}

In the standard approach to the shallow water wave problem, the fluid is assumed to be inviscid and incompressible and the fluid motion to be irrotational.  Therefore a velocity potential $\phi$ is introduced. It satisfies the Laplace equation with appropriate boundary conditions. The Laplace equation must be valid for the whole volume of the fluid, whereas the equations for boundary conditions are valid at the surface of the fluid and at the impenetrable bottom. The system of equations for the velocity potential $\phi(x,y,z,t)$, including its derivation, can be found in many textbooks, for instance, see  \cite[Eqs.\ (5.2a-d)]{Rem}. A standard procedure consists in introducing two small parameters $\alpha=a/H$
and $\beta=(H/L)^2$, where $a$ is a typical amplitude of a surface wave $\eta$, $H$ is the depth of the container and $L$ is a typical wavelength of the surface waves. The parameters  $\alpha,\beta$ are the same as the parameters $\varepsilon,\delta^2$ in \cite{Rem}, respectively. In these notations we follow the paper \cite{B&S}, where a systematic way for the derivation of wave equations of different orders is presented. In \cite{KRR} we introduced a third parameter $\delta=a_h/H$, where $a_h$ is the amplitude of bottom variation. With this new parameter we are able to consider the motion of surface waves over a non-flat bottom within the same perturbative approach as for derivation of KdV or higher-order KdV-like equations. 

\begin{figure}[tb]
\begin{center}
 \resizebox{0.999\columnwidth}{!}{\includegraphics{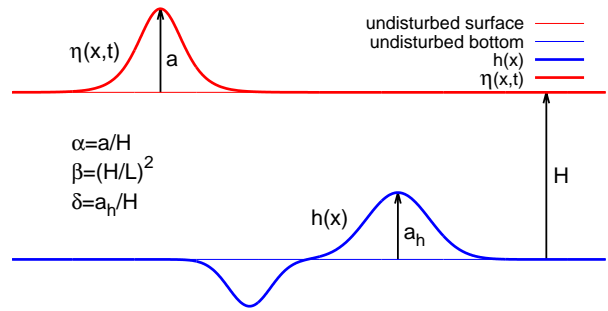}}
 \caption{Schematic view of the geometry of the shallow water wave problem.} \label{geom}
\end{center} \label{F1}
 \end{figure}

In the following we limit our considerations to  2-dimensional flow, $\phi(x,z,t), \eta(x,t)$, where $x$ is the horizontal coordinate and $z$ is the vertical one (this means translational symmetry with respect to $y$ axis).  The geometry of the problem is sketched in Fig.~\ref{geom}. 

Up to now, a generally small surface tension term has been neglected, but   it can be taken into account. A third coordinate could also be included \cite {InR}.

 It is convenient to study the problem in non-dimensional variables.  
The non-dimensional variables are defined as follows  
\begin{eqnarray} \label{bezw}
 \tilde{\eta} & =&
\eta/a,\quad \tilde{\phi}= \phi /(L\frac{a}{H}\sqrt{gH}),\quad \tilde{h} = h/H,\nonumber\\  
\tilde{x}  & =&
x/L,\quad \tilde{z}= z/H, \quad \tilde{t}= t/(L/\sqrt{gH}). 
\end{eqnarray}
 
In the non-dimensional variables the set of hydrodynamic equations for 2-dimensional flow takes the following form (henceforth all tildes have been omitted)
\begin{eqnarray} \label{2BS}
\beta \phi_{xx}+\phi_{zz} & = &0, \\ \label{4BS}
\eta_t+\alpha\phi_x\eta_x-\frac{1}{\beta}\phi_z & = &0,~\mbox{for}~  z = 1+\alpha \eta\\  \label{5BS}
\phi_t+\frac{1}{2}\alpha \phi_x^2+\frac{1}{2}\frac{\alpha}{\beta}\phi_z^2 +\eta 
& = &0,~\mbox{for}~  z = 1+\alpha \eta \\ \label{6BS}
\phi_z-\beta\delta\left( h_x\,\phi_x\right) & = & 0,~\mbox{for}~ z=\delta h(x) .
\end{eqnarray}
Equation (\ref{2BS}) is the Laplace equation, valid for the whole volume of the fluid. Equations (\ref{4BS}) and (\ref{5BS}) are so called kinematic and dynamic boundary conditions at the surface, respectively. Equation  (\ref{6BS}) represents a boundary condition at the non-flat  bottom. All subscripts denote  partial derivatives with respect to particular variables, i.e.\ $\phi_{xx}\equiv \frac{\partial^2 \phi}{\partial x^2}$ and so on. 

For the standard KdV case, the boundary condition at the bottom is $\phi_z=0$.
When the bottom varies, this condition (in original variables) has to be replaced by $ \phi_z=h_x\,\phi_x$, which in non-dimensional variables takes the form (\ref{6BS}). However, in order to ensure that the perturbative approach makes sense, we assume that nowhere are derivatives of $h(x)$ very large.

\section{Derivation of the second order wave equation} \label{deriv}

The details of the derivation of the  nolinear wave equation for the function $\eta(x,z,t)$ when the bottom is given by an arbitrary function $h(x)$ are presented in a previous paper \cite{KRR}. 
In order to make this paper self-contained, the main points of that derivation are recalled here. The present derivation differs from the previous one because here all second order corrections are obtained in one step.
However, the emphasis will be on some surprising features of the evolution of solitary waves governed by this  equation as obtained by numerical simulations.

As in the standard approach, the velocity potential is approximated in the form of the series
$\phi(x,z,t)=\sum_{m=0}^\infty z^m\, \phi^{(m)} (x,t).$
In our derivation (as in most) the velocity potential is limited to a polynomial with $m\le 6$ and in the equations (\ref{2BS})-(\ref{6BS}) only terms up to  second order in small parameters $\alpha,\beta,\delta$ are retained.
Laplace equation (\ref{2BS}) allows us to express all $\phi^{(2m)}$ functions  by the derivatives $\phi^{(0)}_{2mx}$ and $\phi^{(2m+1)}$ functions by the derivatives $\phi^{(1)}_{2mx}$. 

Limiting the  boundary condition at the bottom (\ref{6BS}) to the second order in small parameters, i.e.\ to
\begin{equation} \label{ord1}
\phi^{(1)}(x,t)= \beta\delta \left(h_x \phi^{(0)}_{x} +h  \phi^{(0)}_{2x}\right),
\end{equation}
one is able to express all functions $\phi^{(m)}$  by  $\phi^{(0)}, h $ and their derivatives. (Next term in (\ref{6BS}) is of the order of $\beta\delta^2$. Its inclusion introduces a difficult differential equation for $\phi^{(1)}(x,t)$).
The resulting velocity potential is 
\begin{eqnarray} \label{pot1} \hspace{-4ex}\phi   &\! =\!  &    \phi^{(0)}\!+\!z\beta\delta \left(\! h  \phi^{(0)}_x\! \right)_{x}\!\! -\!\frac{1}{2}z^2 \beta \, \phi^{(0)}_{2x}  \!-\!\frac{1}{6}z^3 \beta^2\delta \left(\! h\phi^{(0)}_x \! \right)_{3x} \nonumber  \\ \!&\!\!& \! +\! \frac{1}{24}z^4 \beta^2 \phi^{(0)}_{4x}  \!+\! \frac{1}{120}z^5 \beta^3\delta\left(\! h \phi^{(0)}_x \! \right)_{5x} \! \!+\!  \frac{1}{720}z^6 \beta^3 \phi^{(0)}_{6x}.  \end{eqnarray}
In the next steps we insert  $\phi(x,z,t)$ given by (\ref{pot1}) into (\ref{4BS}) and (\ref{5BS}), then neglect terms of order  higher than  second  in small parameters $\alpha, \beta, \delta$. Equation (\ref{5BS}) is then differentiated with respect to $x$ and $w(x,t)$ is substituted in place of $\phi^{(0)}_{x}(x,t)$ in both equations. In this way a set of two coupled nonlinear differential equations is obtained which, in general, can be considered at different orders of the approximation.

Keeping only terms up to second order (to be consistent with the order of approximation used in the bottom boundary condition) one arrives at the second order Boussinesq system
\begin{eqnarray} \label{4hx}
0 \!\!&=&\!\!  \eta_t + w_x +   \alpha(\eta w)_x-\frac{1}{6}\beta w_{3x}-\frac{1}{2} \alpha\beta (\eta w_{2x})_x \nonumber  \\
\! &\! \!&\! \hspace{7.5ex}  + \frac{1}{120}\beta^2 w_{5x} -
 \delta (hw)_x +\frac{1}{2}\beta\delta (hw)_{3x}  \\ \label{5hx}
0 \!\!&=&\!\! 
w_t + \eta_x + \alpha w w_x -\frac{1}{2}\beta\, w_{2xt} + \frac{1}{24}\beta^2\, w_{4xt} + \beta\delta\, (h w_t)_{2x}  \nonumber  \\
 &&  \hspace{7.5ex} 
+ \frac{1}{2} \alpha\beta\left[-2(\eta w_{xt})_x +  w_x w_{2x} - w w_{3x} \right] . 
\end{eqnarray}
In (\ref{4hx}) there are two terms depending on the variable bottom, 
the first order term $ \delta (hw)_x$ and the second order term $\frac{1}{2}\beta\delta (hw)_{3x}$, whereas  (\ref{5hx}) contains only the second order term  $\beta\delta (h w_t)_{2x}$. However, the bottom boundary condition (\ref{ord1}), which is the source of these terms, is already  second order in $\beta\delta$. Therefore we will treat all these terms on the same footing, as  second order ones, i.e.\ replacing $ \delta\, (hw)_x$ by $ \beta\delta\, (hw)_x/b,~ b\ne 0$, during derivations and substituting $b=\beta$ in the final formulas. So, we consider equation (\ref{4hx}) in a slightly reformulated form
\begin{eqnarray} \label{4hxa}
\eta_t + w_x \! & +&\!    \alpha\,(\eta w)_x-\frac{1}{6}\beta\, w_{3x} 
\! -\!
 \frac{1}{2} \alpha\beta \,(\eta w_{2x})_x \!+\!\frac{1}{120}\beta^2\, w_{5x} \nonumber \\ 
\! &\! +\!&\!\frac{1}{2}
 \beta\delta\left( -\frac{2}{b}(hw)_x +(hw)_{3x}\right)=0 .
\end{eqnarray}

It is now time to eliminate one of the variables, that is $w(x,t)$, in order to obtain a single equation for the wave shape $\eta(x,t)$.  
Substituting $\delta=0$ (i.e.\ flat bottom) and keeping only first order terms one easily obtains the KdV solution (see, e.g.\ \cite[App.\ C]{Rem} or \cite[eqs.~(13)-(18)]{B&S}). 
Burde and Sergyeyev \cite{B&S} have shown how to proceed with approximations of higher order, assuming the case of the flat bottom. In our paper \cite{KRR} the method of incorporating a variable bottom in the second order perturbative approach is presented .

Burde and Sergyeyev \cite{B&S} showed how to eliminate sequentially the 
$w(x,t)$ variable and obtain a single equation for $\eta(x,t)$ for the higher order perturbative approach. Their method consists in applying special properties of solutions to lower order equations for $w$ and $\eta$ in derivations of corrections to equations in the next order. In principle it can be applied up to an arbitrary order.

In order to obtain a single equation for the elevation function, we take the seond order trial function $w(x,t)$ in the following form
\begin{eqnarray} \label{wab2}
w(x,t) \!& = &\! \eta-\!\frac{1}{4}\alpha\,\eta^2 +\!\frac{1}{3}\beta\,\eta_{2x} 
 + \!
\alpha^2\, \mbox{Q}\alpha^2(x,t) +\!\beta^2 \, \mbox{Q}\beta^2(x,t) \nonumber \\
\!& + &\!
\alpha\beta\, \mbox{Q}\alpha\beta(x,t) + \beta\delta\mbox{Q}\beta\delta(x,t),
\end{eqnarray}
where $\mbox{Q}\alpha^2, \mbox{Q}\beta^2, \mbox{Q}\alpha\beta, \mbox{Q}\beta\delta$ are unknown functions of $\eta, h$ and their derivatives. Insertion of the trial function (\ref{wab2}) into (\ref{5hx}) and  (\ref{4hxa}), use of the properties of the first order equation
\begin{equation} \label{1rz}
\eta_t= -\eta_x-\frac{3}{2}\alpha\,\eta\eta_x -\frac{1}{6}\beta\,\eta_{3x}
\end{equation} 
and rejection of higher order terms, 
yields a set of two equations containing derivatives of unknown functions.
Both of them contain only  second order terms, as lower order terms cancel .
Then we substract these equations. Because we can treat small parameters as independent of each other, the coefficients in front of $\alpha^2, \beta^2, \alpha\beta, \beta\delta$  vanish sparately. This procedure gives 
\begin{eqnarray} \label{rr}
-\mbox{Q}\alpha^2_t+\mbox{Q}\alpha^2_x-\frac{3}{4} \eta ^2 \eta_x & = & 0, \\ \label{r1}
-\mbox{Q}\beta^2_t +\mbox{Q}\beta^2_x-\frac{1}{5} \eta_{5x}& = & 0, \\ \label{r2}
-\mbox{Q}\alpha\beta_t +\mbox{Q}\alpha\beta_t-\frac{7}{4} \eta_x \eta_{2x}- \eta \eta_{3x} & = & 0, \\ \label{r3}
-\mbox{Q}\beta\delta_t(x,t) +\mbox{Q}\beta\delta_x(x,t) -\frac{(h\eta)_x}{b}+\frac{1}{2}h_{3x}\eta  && \\ 
+ \frac{5}{2}h_{2x}\eta_x +\frac{7}{2} h_x\eta_{2x} +\frac{3}{2}h\eta_{3x} &= & 0.\nonumber
\end{eqnarray}
Because the correction functions appear already in the second order, it is enough to use the zero order relation between their time and space derivatives. Therefore we use $Q_t=-Q_x$ (like $\eta_t=-\eta_x, w_t=-w_x$) in all equations (\ref{rr})-
 (\ref{r3}), which allows us to integrate these equations and obtain analytic forms of all correction functions. The derivation of the correction term $\mbox{Q}\beta\delta$ presented here differs from that in \cite{KRR}, where corrections $\mbox{Q}\alpha^2,\mbox{Q}\beta^2, \mbox{Q}\alpha\beta$ where calculated first and $\mbox{Q}\beta\delta$ was obtained in the next step. The final result is the same since differences only appear in  third order.

So, finally we obtain the equations (restoring $b=\beta$)
\begin{eqnarray} \label{wwabd}
w &=& \eta-\alpha\frac{1}{4}\eta^2 +\beta\frac{1}{3}\eta_{2x}+\alpha^2\frac{1}{8} \eta^3 +\beta^2 \frac{1}{10} \eta_{4x} \nonumber \\ && \hspace{2ex}
+ \alpha\beta \left( \frac{3}{16} \eta_{x}^2 + \frac{1}{2}\eta \eta_{2x}\right)  \\  && \hspace{2ex}
 +\beta\delta\left(\! \frac{(2h-\beta h_{2x})\eta}{4\beta} - h_x\eta_x - \frac{3}{4}h\eta_{2x}\!\right) \nonumber
\end{eqnarray}
 and
\begin{eqnarray} \label{etaabd} &&  \eta_t+\eta_x + \alpha \frac{3}{2}\eta\eta_x +\beta\frac{1}{6} \eta_{3x} + \alpha^2\left(\!-\frac{3}{8}\eta^2\eta_x \!\right)  \nonumber \\ && \hspace{3ex} + \alpha\beta\left(\!\frac{23}{24}\eta_x\eta_{2x}\!+\!\frac{5}{12}\eta\eta_{3x}\! \right)+\beta^2\frac{19}{360}\eta_{5x}  \\ && \hspace{3ex} +\beta\delta \frac{1}{4}\left(\!-\!\frac{2}{\beta}(h\eta)_x \!+\! (h_{2x}\eta)_x \!-\! (h\eta_{2x})_x\! \right) = 0. \nonumber \end{eqnarray}
The equation (\ref{etaabd}) is possibly the first KdV-type equation containing terms originating from an uneven bottom in the lowest possible order. It is not yet clear whether analytical solutions of  (\ref{etaabd}) for some non flat cases of the bottom function $h(x)$ can be found. 
It does seem that the inverse scattering transform method (IST) \cite{GGKM,Ablc,InR}, so succesfull in the search of analytical solutions to the KdV equation, cannot be  applied to  equation (\ref{etaabd}).
However, numerical solutions, which have also inspired past analytical studies, for some particular initial conditions should be obtained relatively simply.

The KdV equation posesses an infinite number of invariants, see, e.g.\ \cite[Sec.5.1]{DrJ}, that is, functions of $\eta$ which are constants in time. Do similar invariants exist for the second order equation (\ref{etaabd})? Indeed, there obviously is at least one such invariant, $\int_{-\infty}^{\infty}\eta(x,t)dx=const$. To see this property it is enough to transform the equation   (\ref{etaabd}) to the form 
$\frac{\partial}{\partial t}\eta+\frac{\partial}{\partial x}f(\eta,h)=0 $ and integrate over the whole space. For the eq.\  (\ref{etaabd}) the function $f(\eta,h)$ is
 \begin{eqnarray*}
f(\eta,h)&=&\eta+\frac{3}{4}\alpha \eta^2 - \frac{1}{8}\alpha^2\eta^3
+ \alpha\beta\left(\frac{13}{38} \eta_x^2+\frac{5}{12}\eta\eta_{2x}\right) \\ 
&+& \frac{19}{360}\beta^2\eta_{4x} +\beta\delta\left( -\frac{h\eta}{2\beta}
+\frac{1}{4}h_{2x}\eta -\frac{1}{4} h \eta_{2x}\right). 
\end{eqnarray*}  
If limits of $\eta, h$ and their space derivatives are zero or the same constants when $x\to\pm\infty$, then the conservation law \begin{equation} \label{coneq3}
\int_{-\infty}^{\infty}\eta(x,t)\,dx = \mbox{constant,} \end{equation}
holds. It is clear that the same conservation law holds for the case $\delta=0$, i.e., for the second order equation with flat bottom. The existence of other invariants for wave motion described by the second order equation (\ref{etaabd}) is still an open question.
We are looking into it.

\section{Solution to our second order equation for a flat bottom} \label{nsol}

We seek a solution to (\ref{etaabd}) with $\delta=0$, or
\begin{eqnarray} \label{etaab}
\eta_t+\eta_x &+& \alpha\, \frac{3}{2}\eta\eta_x +\beta\,\frac{1}{6} \eta_{3x} +
\alpha^2 \left(-\frac{3}{8}\eta^2\eta_x\right)  \\
&+& \alpha\beta\,\left(\frac{23}{24}\eta_x\eta_{2x}+\frac{5}{12}\eta\eta_{3x} \right)+\beta^2\,\frac{19}{360}\eta_{5x} =0.  \nonumber
\end{eqnarray}

Assume the form of a soliton moving to the right, $\eta(x,t)=\eta(x-vt)$.
 we have, $\eta_t=-v\eta_x$ and (\ref{etaab}) can be written as 
\begin{eqnarray} \label{eta1}
(1-v)\eta_x &+& \alpha\, \frac{3}{2}\eta\eta_x +\beta\,\frac{1}{6} \eta_{3x} -\frac{3}{8}
\alpha^2 \eta^2\eta_x \\
&+& \alpha\beta\,\left(\frac{23}{24}\eta_x\eta_{2x}+\frac{5}{12}\eta\eta_{3x} \right)+\beta^2\,\frac{19}{360}\eta_{5x} =0.  \nonumber
\end{eqnarray}
Integrating, one obtains
\begin{eqnarray} \label{eta2}
(1-v)\eta &+& \alpha\, \frac{3}{4}\eta^2 +\beta\,\frac{1}{6} \eta_{2x} -\frac{1}{8}
\alpha^2 \eta^3 \\
&+& \alpha\beta\,\left(\frac{13}{48}\eta_x^2+\frac{5}{12}\eta\eta_{2x} \right)+\beta^2\,\frac{19}{360}\eta_{4x} =0.  \nonumber
\end{eqnarray}

We look for a solution $\eta(x,t)=A\,\mbox{Sech}^2\left(B(x-vt)\right)\equiv A\,\mbox{Sech}^2(By),~ y=x-vt  $.

Using (\ref{eta2}) and the properties of $\mbox{Sech}^2$ we obtain
\begin{equation} \label{war1}
C2\, \mbox{Sech}^2(By) + C4\, \mbox{Sech}^4(By) +C6\, \mbox{Sech}^6(By)=0,
\end{equation}
where (upon dividing (\ref{war1}) by ~$A$)
\begin{eqnarray} \label{c2}
C2 &=& (1-v) + \frac{2}{3}  B^2 \beta + \frac{38}{45}  B^4 \beta^2 \\ \label{c4}
C4 &=&  \frac{3 A \alpha}{4} -  B^2 \beta + \frac{11}{4} A\alpha\, B^2  \beta -  \frac{19}{3} B^4 \beta^2\\ \label{c6}
C6 &=&  -\left(\frac{1}{8}\right) (A \alpha)^2 - \frac{43}{12} A\alpha\, B^2  \beta +  \frac{19}{3}  B^4 \beta^2
\end{eqnarray}
From (\ref{c6}), denoting $\displaystyle z=\frac{\beta B^2}{\alpha A}$ we obtain
\begin{equation} \label{rkw}
\frac{19}{3} z^2 - \frac{43}{12} z -\frac{1}{8}=0,
\end{equation}
solved by
\begin{equation} \label{rkw1}
\begin{array}{lll} z_1 & = &\displaystyle \frac{43-\sqrt{2305}}{152}\approx -0.033<0\\
 z_2 & = & \displaystyle \frac{43+\sqrt{2305}}{152}\approx 0.6 > 0. \end{array}
\end{equation}

Thus
\begin{equation} \label{wB}
\left(\frac{B^2}{A}\right)_{1/2} = \frac{\alpha}{\beta} z_{1/2},
\end{equation}
with ~$A< 0$~ for ~$z=z_1< 0$~ and ~$A> 0$~ for ~$z=z_2> 0$.

Inserting ~$\beta B^2=\alpha A \,z$~ into (\ref{c4}) we have:
\begin{equation}\label{a4}
A=\frac{z-\frac{3}{4}}{\alpha\,z (\frac{11}{4}-\frac{19}{3}z)}
\end{equation}
therefore for ~$z=z_1\approx -0.033$~  follows ~$A>0$, leading to trouble 
(\ref{wB}). 

However, for ~$z=z_2\approx 0.6$~ we obtain $A>0$, no contradiction with (\ref{wB}). Thus we have exactly one solution of (\ref{rkw})
~~$\displaystyle z= z_2=\frac{43+\sqrt{2305}}{152}\approx 0.598752733793626.$

Since ~$\beta B^2=\alpha A \,z$,~ so \begin{equation} \label{bb}
B = \sqrt{ \frac{z-\frac{3}{4}}{\beta (\frac{11}{4}-\frac{19}{3}z)}}
\end{equation}

\vspace{2mm}
Now from (\ref{c2}) we obtain
\begin{equation}\label{v1}
v = 1 + \beta B^2(\frac{2}{3}+\frac{38}{45}\beta B^2 ).
\end{equation}
Using (\ref{a4}) and  $\beta B^2= \alpha\,A\,z =
\displaystyle \frac{z-\frac{3}{4}}{(\frac{11}{4}-\frac{19}{3}z)}$~ yields 
\begin{equation}\label{v2}
v = 1 + \frac{z-\frac{3}{4}}{(\frac{11}{4}-\frac{19}{3}z)}\left(\frac{2}{3}+\frac{38}{45} \frac{z-\frac{3}{4}}{(\frac{11}{4}-\frac{19}{3}z)} \right) \approx 1.114546.
\end{equation}

We have found the single-soliton solution to the second order equation (\ref{etaab})
\begin{equation}\label{sol}
\eta(x,t)=A\,\mbox{Sech}^2\left[B\left(x-v\,t\right)\right],
\end{equation}
 for which ~$A$,~$B$~ and ~$v$~ are given by (\ref{a4}), (\ref{bb}) and (\ref{v2}). We will call this solution the {\em second order KdV soliton, in abbreviation KdVII soliton}.

The soliton (\ref{sol}) should satisfy (\ref{etaab}). This is confirmed by numerics, see fig.~\ref{F8}.

It is worth to emphasize, that contrary to the claim in \cite{SK_PO}, cited in the Introduction, there exists an analytic solution to the second order KdV-type equation (\ref{etaab}).

\section{Numerical studies} \label{numer}

In our previous paper \cite{KRR} the first examples of numerical calculations for the time evolution of a KdV soliton  according to the second order equation (\ref{etaabd}) were presented. However, the examples for a non-flat bottom were limited to short time evolution. In this paper we have focused on much longer times. 

\subsection{Initial condition in the form of KdV soliton} \label{sub1}

 All the calculations presented below are  in non-dimensional variables (\ref{bezw}). In all examples presented in this subsection we assume the initial wave as the exact single KdV soliton 
$\eta(x,t)= \mbox{sech}\left[\frac{\sqrt{3}}{2}\left(x-x_0-t(1+\frac{\alpha}{2})\right)\right]^2$
 at $x_0=0$, $t=0$ (in non-dimensional variables we took the amplitude of the soliton to be 1). The algorithm used was the Zabusky-Kruskal one \cite{ZK}, modified in order to include terms of second order . The space derivatives of $\eta(x,t)$ were calculated numerically step by step from the grid values of the function and lower order derivatives by a nine-point central difference formula.
Calculations were performed on the interval $x\in[0,D]$ with the periodic boundary conditions of $N$ grid points. The space grid points were separated by $\Delta x=0.05$. The time step $\Delta t$ was chosen as in \cite{ZK}, i.e.,  $\Delta t=(\Delta x)^3/4$. The calculations shown in this paper used grids with $N=4400$ and $N=13200$, implying $D=220$ and $D=660$. 
For the soliton motion covering the interval $x\in [0,D]$ the number of time steps reaches 2$\cdot 10^7$. In all cases the algorithm secures the volume (mass) conservation (\ref{coneq3}) up to 8-10 decimal digits.  The initial position of the soliton is $x_0=0$ in all  cases.  

\begin{figure}[tb]
 \resizebox{0.999\columnwidth}{!}{\includegraphics{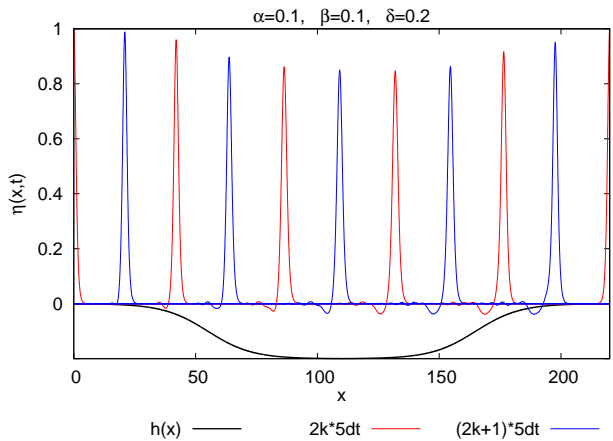}}
\caption{Time evolution of the initial KdV soliton according to Eq.~(\ref{etaabd}) for bottom shape function $h_-(x)$. See detailed explanations in the text.} \label{F2}\end{figure}

\begin{figure}[b]
 \resizebox{0.999\columnwidth}{!}{\includegraphics{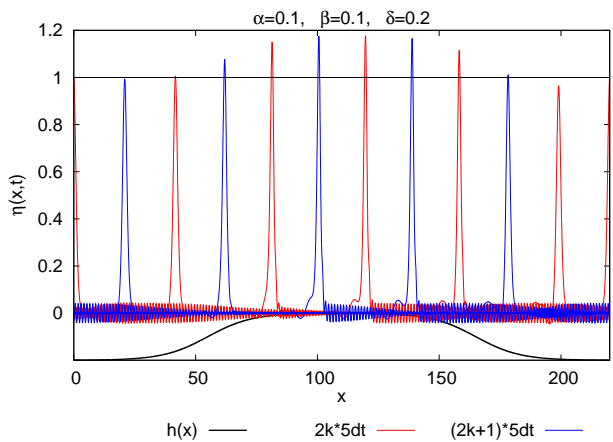}}
\caption{The same as in Fig.~\ref{F2} but for the bottom shape function $h_+(x)$.}
\label{F3}
\end{figure}

We begin calculations with the bottom function defined as  $h_{\pm}(x)=\pm\frac{1}{2}[\mbox{tanh}(0.055(x-55))+1]$ for $x\le 110$ and its symmetric reflection with respect to $x=110$ for $x>110$.
Fig.~\ref{F2} presents snapshots of the time evolution of the initial wave, according to  Eq.~(\ref{etaabd}), over the bottom, defined by $h_-(x)$ function. The red curves show the shapes of the wave at  time instants $t_i= 0,10,20,30,40*dt$, where $dt=4$, whereas the blue ones correspond to times $t_i= 5,15,25,35*dt$. The same color scheme is used in the next figures. 
One observes a decrease in the amplitude of the wave when the depth of water increases and the inverse behavior when the bottom slants up. The small backscattered tail increases slowly with time.

In Fig.~\ref{F3} the same sequence of snapshots for the soliton motion is presented for the bottom function $h_+(x)$. Here one observes at first an increase then a decrease in the amplitude of the main wave. In the case when the main part of the wave approaches a shallower region a forward scattering occurs and creates waves of much smaller amplitude outrunning the main one.

A closer inspection of the results presented above brings to light interesting relations between the bottom changes and amplitude and velocity of the main wave.
 When the pure KdV equation is considered (corresponding to a limitation of  Eq.~(\ref{etaabd}) to  first order and flat bottom) the amplitude of the soliton and its velocity is greater when the water depth is smaller. Therefore, from this point of view, one expects a slower soliton motion when it enters a deeper basin and a faster motion when it moves towards a shallowing.  
On the other hand,  inspection of solutions to the KdV-type equation obtained in \cite{G&P1}, (see, e.g.\ Figs.~3 and 4), which is  second order in the small parameter for slow bottom changes, shows qualitatively that when the depth  decreases, the amplitude of the solitary wave increases with simulatneous a decrease of its wavelength and velocity. (The small paremeter used in \cite{G&P1,G&P2} is different than ours, as it measures the ratio of the bottom variation to a wavelength.) A decrease of the velocity with simultaneous increase of the amplitude (and a creation of slower secondary waves) is obtained for the solitary wave entering a shallower region in \cite[see, Fig.1]{CV_DD}, as well.

The distances between the peaks  shown in Figs.~\ref{F2} and \ref{F3} indicate that the main waves in Fig.~\ref{F2} cover, in the same time periods,  larger distances over a deeper water than the waves in Fig.~\ref{F3} travelling over  shallower water. The corresponding sequence od decrease/increase of the wave's amplitude in Fig.~\ref{F2} and increase/decrease in Fig.~\ref{F3} is clearly visible.

\begin{figure}[t]
 \resizebox{0.999\columnwidth}{!}{\includegraphics{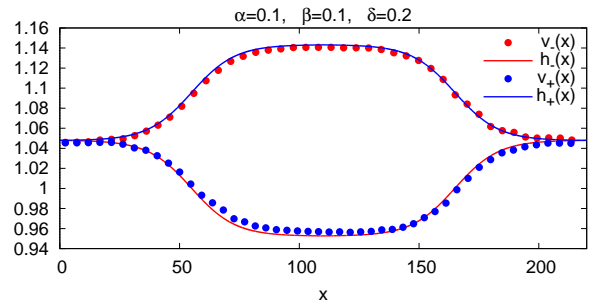}} 
\caption{Anticorrelations between the soliton's velocity and the water depth. Dots indicate the average velocities of the tops of solitons for given positions, lines with the same color the shape of the bottom function.} \label{F4} \end{figure}

\begin{figure}[t]
 \resizebox{0.999\columnwidth}{!}{\includegraphics{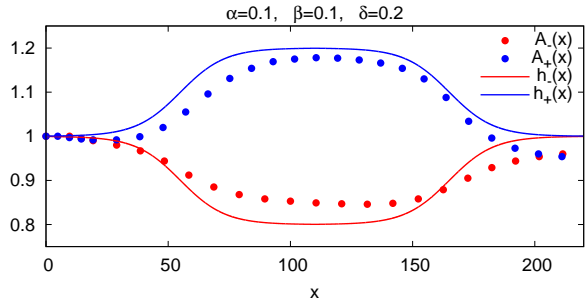}} 
\caption{Correlations between the soliton's amplitude and the water depth. Dots indicate amplitudes of solitons for given positions, lines with the same color the shape of the bottom function.} \label{F4a} \end{figure}

Can we get more precise information on these velocities from our numerical data?
Having recorded the shapes of solitons $\eta(x,t_k)$ in smaller time steps than those presented in Figs.~\ref{F2}-\ref{F3}, we made an effort to estimate the average values of the velocities for a given time step. Define 
\begin{equation} \label{vel} v(x,t_i)=\frac{X(t_i)-X(t_{i-1})}{t_i-t_{i-1}},
\end{equation}
where $X(t_i)$ is the position of the {\bf top} of the wave. Because this position, due to the finite space grid, is read off by interpolation, the values of $X(t_i)$ have precision limited to 4-5 digits . This is enough, however, to observe an almost perfect anticorrelation of these velocities with the depth. Contrary to "obvious" conclusions from KdV reasoning, Fig.~\ref{F4} shows that
when the water depth increases, the average velocity of the top of the wave likewise increases and vice versa. From plots of the bottom functions $h(x)$, appropriately scaled and vertically shifted, one sees that this correlation is almost linear. Concerning numerical values, note that the velocity of the KdV soliton is $v_{KdV}=1+\frac{\alpha}{2}=1.05$. 
Similar, however less linear, correlations occur between the water depth and the soliton's' amplitude. It is presented in Fig.~\ref{F4a}.

\begin{figure}[tbh]
 \resizebox{0.999\columnwidth}{!}{\includegraphics{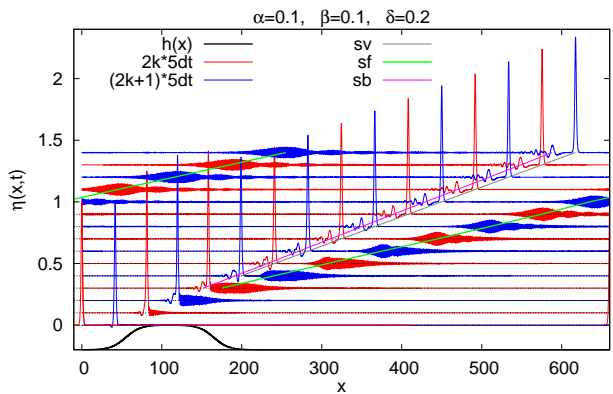}} 
\caption{Distortions of solitary wave due to the motion over an extended obstacle. See details in the text.}
\label{F5} \end{figure}

The forward scattered waves seen in Fig.~\ref{F3}  suggest that something interesting can occur at later stages of the wave motion. However, in order to eliminate the influence of "neighbor cell effects" arising from the periodic boundary conditions, we decided to check this with an interval three times longer, $x\in [0:660]$ in which the bottom varies only in the first part of that interval. Several snapshots of the wave motion in that setting are shown in Fig.~\ref{F5}. In this case, the calculated data are plotted at time steps of $2k\cdot dt$ and $(2k+1)dt$, $k=,1,\ldots,7$,  where $dt=8$. Comparing waves at time instants ~$t=15 dt, ~20 dt, ~25 dt,\ldots\;$ (where the parts of the waves are
still far from the boundary) one sees sequential formation of the forward wave train in the form of a wave packet. This wave packet comes from the main part (a solitary wave) and  moves faster than the main wave. Then this wave packet divides  at later stages of the motion. The thick green line going through the positions of the top of the envelope of this wave packet indicates the constant velocity of that part of the wave. 
Two other thick lines,  grey  and  magenta , join the  positions of the main soliton and the smaller one, scattered backward, respectively. 
All three lines show the constant (but different) velocities of  these objects when the wave has already passed the obstacle and moves over a flat bottom.

\begin{figure}[t]
 \resizebox{0.999\columnwidth}{!}{\includegraphics{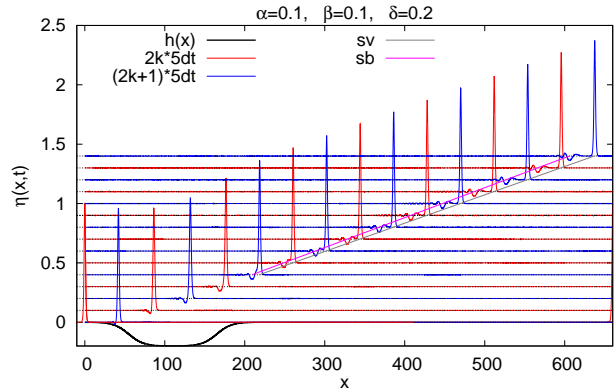}} 
\caption{Distortions of a solitary wave due to motion over an extended well.}
\label{F6} \end{figure}

\begin{figure}[t]
 \centerline{\resizebox{0.53\columnwidth}{!}{\includegraphics{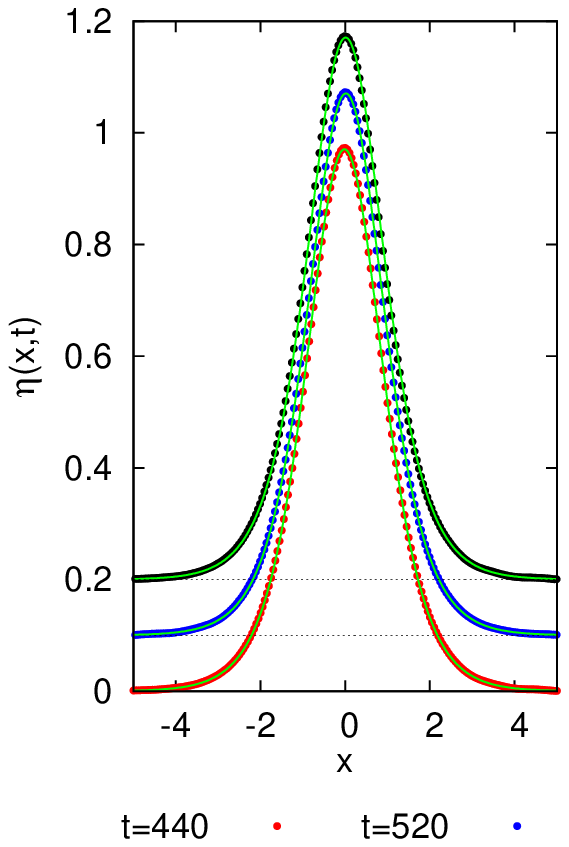}}\hspace{-3mm} \resizebox{0.53\columnwidth}{!}{\includegraphics{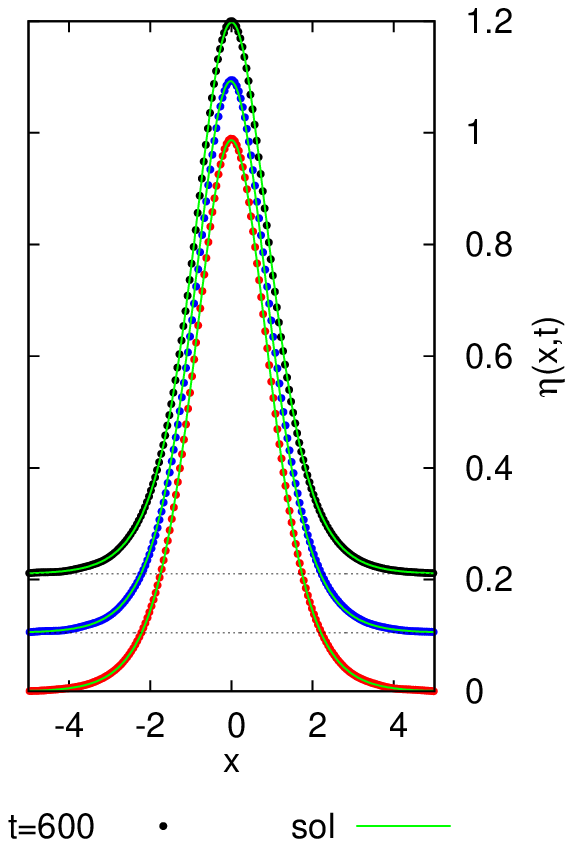}} }
\caption{Comparison of shapes of the main part of the waves at $t=440,520,600$ from Fig.~\ref{F5} (left) and Fig.~\ref{F6} (right) with the shape of the KdV soliton (green line) after shifts to the same position.}
 \label{F7} \end{figure}

Fig.~\ref{F6} shows the long time evolution of the initial soliton above an extended well of the same shape and amplitude as the obstacle in the previous case. Here only one backward scattered wave is seen. Its velocity, indicated by the thick magenta line, is only a little smaller than the velocity of the main part of the wave.
 
Does the main part of the wave preserve the shape of the KdV soliton  when it is moving over the flat bottom region after passing the interval of varying bottom? In order to answer this question we compared the shapes of the main part of the wave at temporal points $t=440,520,600$ with  the shape of KdV soliton.

In Fig.~\ref{F7} the shapes of the main part of the waves after a long period of evolution, shown in Figs.~\ref{F5} and \ref{F6}, are compared with the shape of the KdV soliton. The comparison was made as follows: for each time instant $t_i$, we selected an interval $x\in[x_{top}(t)-5,x_{top}(t)+5]$, where $x_{top}(t)$ was the position of the top of that wave; then we fitted the formula $f(x,t)=a\,\mbox{sech}[b(x-ct)]^2$ to values of $\eta(x,t)$  recorded in grid points as solutions of  Eq.~(\ref{etaabd}). The dots in Fig.~\ref{F7} represent numerical solutions to (\ref{etaabd}), whereas the green lines represent the fitted KdV solitons. It is remarkable that, for the given case, it is the same soliton for all time instants when the wave has already passed the obstacle or a well. In the case when the obstacle forms a bump (Fig.~\ref{F5}) the fitted parameters are: $\;a=0.9367$,  $\;b=0.8073$, $\;c=1.0467$. In the case in  Fig.~\ref{F6} the corresponding set is: $\;a=0.9707$,
  $\;b=0.8206$, $\;c=1.0488$. This means that after formation of smaller waves scattered forward and/or backward during interaction with a bottom obstacle the main part preserves the shape of a KdV soliton, although with  slightly smaller apmlitude, width  and velocity.

\subsection{Initial condition in the form of new KdVII soliton (\ref{sol})} \label{sub2} 

\begin{figure}[bht]
 \resizebox{0.999\columnwidth}{!}{\includegraphics{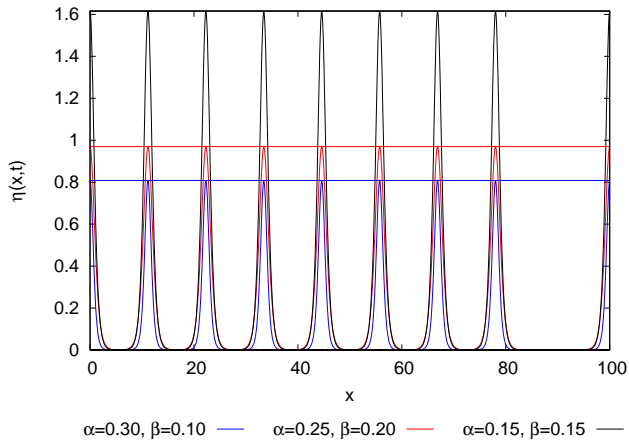}} 
\caption{Time evolution of the exact soliton (\ref{sol}) according to the  the  second order equation (\ref{etaab}) obtained in numerical simulations.}
 \label{F8} \end{figure}

In this subsection we present some examples of the time evolution of the wave which at $t=0$ is given by (\ref{sol}), i.e., it  is the exact solution of the second order KdV-type equation for a flat bottom (\ref{etaab}). In Fig.~\ref{F8} three cases of solitons, corresponding to three different sets of $(\alpha,\beta)$ and moving according to the second order equation (\ref{etaab}) are displayed. In all cases the soliton's velocity is the same, given by (\ref{v2}), what is different from the KdV case, where the velocity depends on $\alpha$. It is clear from the Fig.~\ref{F8} that the numerical solution preserves its shape and amplitude for all cases in agreement with the analytic solution.

\begin{figure}[tbh]
 \resizebox{0.999\columnwidth}{!}{\includegraphics{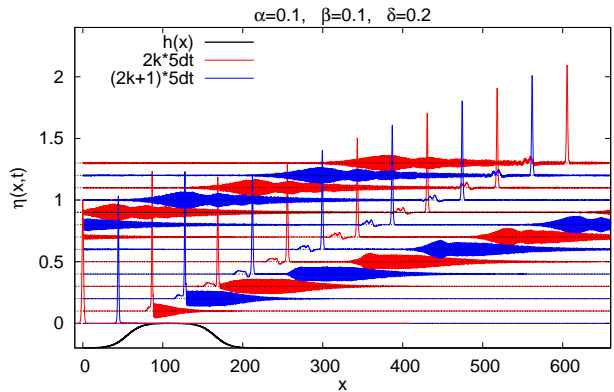}} 
\caption{The same as in Fig.\ \ref{F5} but for initial condition given by the exact second order soliton (\ref{sol}).}
 \label{F9} \end{figure}

\begin{figure}[th]
 \resizebox{0.999\columnwidth}{!}{\includegraphics{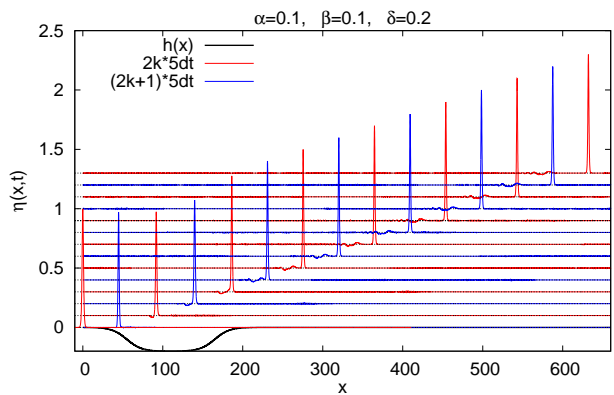}} 
\caption{The same as in Fig.\ \ref{F6} but for initial condition given by the exact second order soliton (\ref{sol}).}
 \label{F10} \end{figure}

In Figs.~(\ref{F9}) and (\ref{F10}) we show the time evolution of the initial soliton (\ref{sol}) according to the equation (\ref{etaabd}) which contains terms from an uneven bottom. In order to compare these cases with the evolution of initial KdV soliton all parameters of the calculations are the same as those related to results shown in Figs.~\ref{F5} and~\ref{F6}. In general the time evolution of initial second order KdV-type soliton  (\ref{sol}) is qualitatively very similar to the evolution of first order soliton (exact KdV soliton). In particular, as seen in Figs.~\ref{F6} and \ref{F10}, time evolution is roughly the same when soliton encounters firstly  deepening and next shallowing of the bottom. There are, however, some differences. First of all the initial velocities of the solitons are slightly different. For exact KdV soliton it is $v_{KdV}=1+\frac{\alpha}{2}=1.05$ for $\alpha=0.1$. Velocity of KdVII soliton (\ref{sol}) does not depend on $\alpha$, $v_{KdVII}\approx 1.114546$.

In cases displayed in Figs.~\ref{F5} and \ref{F9}, when soliton enters firstly   shallowing and then deepening, the wave packet created in front of the KdVII soliton is wider than that in the case of KdV soliton. It moves faster and its fragmentation, in later stages of the evolution, is more pronounced.

In conclusion, we stress that numerical simulations according to the second order KdV-type equation containing terms originating from a varying bottom  (\ref{etaabd}) revealed quantitative results concerning the velocity and amplitude of the solitary wave. 
The initial soliton almost preserves its parameters (shape, amplitude) during the motion over bottom topography being resistant to distortions.


\begin{thebibliography}{99}

\bibitem{kdv} D.J.\  Korteveg and  G.\ de Vries, 
Phil.\ Mag.\ (5), {\bf 39}, 422 (1895). 

\bibitem{Whit} G.B.\ Whitham, {\em Linear and nonlinear waves},  John Wiley \& Sons, New York, (1974). 

\bibitem{DrJ} P.G.\ Drazin and R.S.\   Johnson,  {\em Solitons: An Introduction},  Cambridge University Press, Cambridge, (1989). 
 
\bibitem{InR} E.\  Infeld and G.\ Rowlands,  {\em  Nonlinear Waves, Solitons and Chaos}, Cambridge University Press, Cambridge, (2000), second edition, Chapter 5.

\bibitem{Ablc} M.J.\ Ablowitz, and P.A.\ Clarkson, {\em  Solitons, Nonlinear Evolution Equations and Inverse Scattering}, Cambridge University Press, Cambridge, (1991).

\bibitem{Hir} R.\ Hirota, {\em  The Direct Method in Soliton Theory}, Cambridge University Press, Cambridge, (2004), first published in Japanese (1992).

\bibitem{Rem} M.\ Remoissenet, {\em  Waves Called Solitons: Concepts and Experiments}, Springer, Berlin, (1999).

\bibitem{Mei} C.C.\ Mei and B.\ Le M\'ehaut\'e, 
J.\ Geophys. Research, {\bf 71}, 393-400 (1966).

\bibitem{Grim70} R.\ Grimshaw,  
J.\ Fluid Mech.  {\bf 42}, 639-656 (1970).

\bibitem{Djord} V.D.\ Djordjevi\'c and L.G.\ Redekopp, 
J.\ appl.\ Math. and Phys.\ (ZAMP), {\bf 29}, 950-962 (1978). 
 

\bibitem{BH} E.S.\ Benilov and C.P.\ Howlin, 
Studies in Appl.\ Math., {\bf 116}, 289-301 (2006).

\bibitem{Pel} O.\ Nakoulima, N.\ Zahibo, E.\ Pelinovsky, T.\ Talipova, and A.\ Kurkin, 
Chaos, {\bf 15}, 037107 (2005).

\bibitem{Peli} 
R.\ Grimshaw, E.\ Pelinovsky and T.\ Talipova, 
 Geophys. Astrophys. Fluid Dynamics,  {\bf 102}, 179-194 (2008).

\bibitem{Peli1}
E.\ Pelinovsky, B.H.\ Choi, T.\ Talipova, S.B.\ Woo and D.C.\ Kim, 
 Appl. Math.\ Comput.\,  {\bf 217}, 1704-1718 (2010).

\bibitem{Grim} R.H.J.\ Grimshaw and N.F.\ Smyth, 
J.\ Fluid.\ Mech. {\bf 169} 429-464 (1986).

\bibitem{Smy} N.F.\ Smyth, 
Proc.\ R.\ Soc.\ Lond.\ A,  {\bf 409}, 79-97 (1987).

\bibitem{Kam}
A.M.\ Kamchatnov,  Y.-H.\ Kuo, T.-C.\ Lin, T.-L.\ Horng,  S.-C.\ Gou, R.\ Clift, G.A.\ El, and R.H.J.\ Grimshaw, 
 Phys.\ Rev.\ E {\bf 86},  036605 (2012).
 
\bibitem{G&P1} E.\ van Greoesen and S.R.\ Pudjaprasetya, 
Wave Motion, {\bf 18}, 345-370 (1993).
 
\bibitem{G&P2}  S.R.\ Pudjaprasetya and E.\ van Greoesen, 
Wave Motion, {\bf 23}, 23-38 (1996).
 
\bibitem{GN}
 A.E.\ Green and P.M.\ Naghdi, 
J.\ Fluid Mech., {\bf 78}, 237-246 (1976).

\bibitem{Nad}
B.T.\ Nadiga,  L.G.\ Margolin and P.K.\ Smolarkiewicz, 
Phys. Fluids,   {\bf 8}, 2066-2077 (1996).

\bibitem{Kim}
J.W.\ Kim,  K.J.\ Bai, R.C.\ Ertekin  and W.C., Webster, 
J.\ Eng.\ Math., {\bf 40}, 17-42 (2001).

\bibitem{NiuYu} X.\ Niu and X.\ Yu,  
Coastal Engineering, {\bf 58}, 143-150 (2011);~
H-W.\ Liu and J-J.\ Xie, 
Coastal Engineering, {\bf 58},) 948-952 (2011).

\bibitem{MULTI} MULTIWAVE PROJECT @ 2012 University College Dublin, http://www.ercmultiwave.eu 

\bibitem{CV_DD}  C.\ Viotti, D.\ Dutykh and F.\ Dias, 
Procedia IUTAM {\bf 11}, 110–118 (2014). 

\bibitem{KRR}
A.\ Karczewska, P.\ Rozmej and \L.\ Rutkowski, 
Phys.\ Scr.\ {\bf 89} (2014) 054026. 

\bibitem{SK_PO} S.\ Kichenassamy and P.\ Olver, 
SIAM J.\ Math.\ Anal., {\bf 23}, 1141-1166 (1992). 

\bibitem{Burde}  G.I.\ Burde, 
Commun.\ Nonlinear Sci.\ Numerical Simulat.\ {\bf 16}, 1314-1328 (2011). 


\bibitem{B&S} G.I.\ Burde and A.\ Sergyeyev, 
J.\ Phys.\ A:\ Math.\ Theor. {\bf 46}, 075501 (2013).

\bibitem{GGKM} 
C.S.\ Gardner, J.M.\ Greene, M.D.\ Kruskal, and R.M. Miura, 
Phys.\ Rev.\ Lett.\ {\bf 19} 1095-1097 (1967).


\bibitem{ZK} N.J.\ Zabusky and  M.D.\ Kruskal, 
Phys.\ Rev.\ Lett.\ {\bf 15}, 240-243 (1965).

\bibitem{Zab}
 N.J.\ Zabusky, 
Phys.\ Rev.\ {\bf 168}, 124-128 (1968).

\end{thebibliography}
\end{document}